\documentclass[conference]{IEEEtran}
\IEEEoverridecommandlockouts
\usepackage{cite}
\usepackage{amsmath,amssymb,amsfonts}
\usepackage{algorithmic}
\usepackage{graphicx}
\usepackage{textcomp}
\usepackage{xcolor}
\def\BibTeX{{\rm B\kern-.05em{\sc i\kern-.025em b}\kern-.08em
    T\kern-.1667em\lower.7ex\hbox{E}\kern-.125emX}}

\usepackage[utf8]{inputenc}
\usepackage[T1]{fontenc}
\usepackage{amsmath,amssymb,amsthm}
\usepackage{booktabs}
\usepackage{array}
\usepackage{cite}
\usepackage{graphicx}
\usepackage[caption=false,font=footnotesize]{subfig}
\usepackage{xcolor}
\newcommand{\A}{\mathcal{A}}
\newcommand{\Lset}{\mathcal{L}}
\newcommand{\Xset}{\mathcal{X}}
\newcommand{\dq}{\Delta q}
\newcommand{\bax}{\beta}
\newcommand{\nz}{n_{0}}
\newcommand{\pz}{p_{0}}
\newcommand{\nA}{n_{\mathrm{A}}}
\newcommand{\pA}{p_{\mathrm{A}}}
\newcommand{\nF}{n_{\mathrm{F}}}
\newcommand{\pF}{p_{\mathrm{F}}}
\newcommand{\nG}{n_{\mathrm{G}}}
\newcommand{\kA}{\kappa_{\mathrm{A}}}
\newcommand{\nK}{\nG(\kA)}   % the Gray crossing of the block that holds the anchored index
\newcommand{\bb}{b}
\newcommand{\gd}{g}

\begin{document}

\title{Box-Shift: Fixed-Complexity Counter-Hypothesis Generation for Soft-Output MIMO Box Decoding}

\thanks{This work has been submitted to the IEEE for possible publication.
Copyright may be transferred without notice, after which this version may
no longer be accessible.}%

\author{\IEEEauthorblockN{Shengchun Yang\IEEEauthorrefmark{1},
Amit Sravan Bora\IEEEauthorrefmark{2},
Emil Matus\IEEEauthorrefmark{1}, and
Gerhard Fettweis\IEEEauthorrefmark{1}}
\IEEEauthorblockA{\IEEEauthorrefmark{1}Vodafone Chair Mobile Communications
Systems, Technische Universit\"at Dresden, Dresden, Germany\\
\IEEEauthorrefmark{2}G.\,S.\ Sanyal School of Telecommunications, Indian
Institute of Technology Kharagpur, Kharagpur, India\\
Email: \{shengchun.yang, emil.matus, gerhard.fettweis\}@tu-dresden.de,
amitbora@gssst.iitkgp.ac.in}}

\maketitle

\begin{abstract}
Box Decoding is a sort-free tree-search multiple-input multiple-output (MIMO) detector with a fixed per-layer cost independent of the quadrature amplitude modulation (QAM) order. Its regular datapath is well suited to hardware implementation, but the detector has so far been limited to hard-output operation in uncoded systems. Its fixed four-point box spans two adjacent Gray-coded QAM levels per axis, leaving most bits without counter-hypotheses for higher-order QAM. This missing counter-hypothesis (MCH) problem prevents the direct computation of finite max-log log-likelihood ratios (LLRs). We propose the \emph{Box-Shift} algorithm, which shifts the existing box to the nearest node carrying the complementary value of each Gray-coded bit at each detection layer. Unlike Bit-Flip, Box-Shift analytically identifies the partially optimal counter-hypothesis using one inverter per bit and updates the path metric through an index-domain increment. The proposed method preserves the sort-free, QAM-order-independent structure of Box Decoding. Coded $4\times4$ MIMO simulations using 5G~NR low-density parity-check (LDPC) coding show that Box Decoding with Box-Shift achieves a $0.7$--$1.5$\,dB gain in $E_b/N_0$ over the same detector with Bit-Flip at a bit error rate of $10^{-5}$, with equal soft-candidate budgets. For the soft-candidate metrics, the incremental update of Box-Shift needs $75\%$ fewer real multiplications than the explicit evaluation of Bit-Flip at a comparable number of real additions.
\end{abstract}

\begin{IEEEkeywords}
MIMO detection, soft-output detection, Box Decoding, counter-hypothesis generation, Gray-coded QAM.
\end{IEEEkeywords}

%=============================================================================
\section{Introduction}\label{sec:intro}
%=============================================================================
Spatial multiplexing and high-order quadrature amplitude modulation (QAM) increase the spectral efficiency of multiple-input multiple-output (MIMO) systems, but also make data detection more demanding. Exhaustive maximum-likelihood (ML) detection searches all $|\mathcal{A}|^N$ symbol vectors for $N$ streams and a constellation $\mathcal{A}$, and its cost grows rapidly with the stream count and the QAM order~\cite{MIMO_Larsson}. Linear zero-forcing (ZF) and linear minimum mean-square-error (LMMSE) detectors have low complexity, but their performance deteriorates in ill-conditioned channels and in the presence of multi-user interference~\cite{MIMO_detection_Hanzo}. Modern systems therefore need low-complexity nonlinear detectors that approach ML performance. 3GPP also defines the reduced-complexity ML (R-ML) receiver as an advanced receiver type for these conditions~\cite{TS38101_4}. \par 

Tree-search-based detectors such as sphere decoding (SD) attain ML performance with channel-dependent, variable complexity~\cite{Viterbo1999SD}, whereas $K$-Best offers fixed complexity but requires a sorting stage at each layer~\cite{Guo2006KBest}. Sort-free variants avoid this sorting stage but still rely on QAM-dependent enumeration~\cite{Mondal2010DKB}. Box Decoding instead expands a fixed `box' around a zero-forcing reference at each layer. It is sort-free and QAM-order-independent and performs close to $K$-Best~\cite{box}. Its variants control tree growth with the MIMO dimension. $K$-Box reintroduces global sorting~\cite{KBox}, probabilistic tree pruning applies a statistical threshold~\cite{box_ptp}, and deterministic pruning selects surviving candidates using a few sign comparisons, preserving a parallel, feed-forward structure~\cite{box_scp}. In this paper, we extend the existing Box Decoding framework to soft-output detection by incorporating bit-level log-likelihood ratio (LLR) computation for coded systems.\par 

List-based soft-output detectors approximate the \emph{a posteriori probability} (APP) LLRs of coded bits using the max-log approximation over a reduced candidate list. Under the max-log approximation, each LLR is the difference between the best metric for one bit value and that for the complementary value, so the list must contain a candidate for each value. However, all candidates in a restricted list may share the same value at a particular bit position, leaving no candidate with the complementary value, i.e., no counter-hypothesis. Then one maximization runs over an empty set and the list yields no finite LLR. This is known as the \emph{missing counter-hypothesis} (MCH), or bit-vacancy, problem~\cite{Chen2025FSD}. \par 

Three families of solutions address this problem. In~\cite{Barbero2008LFSD}, \emph{Extra candidate lists} extend the search until both values of every bit are covered, at the cost of many redundant paths. \emph{LLR clipping} replaces the missing metric with a saturation value~\cite{deJong2005ITS,Milliner2008CSILLR,Shao2026CAHSF}. Although it incurs negligible computational overhead, it biases the LLRs and degrades coded performance. \emph{Bit-Flip} instead generates one counter-hypothesis per bit by inverting that bit in the Gray-coded label of the best estimate~\cite{Liu2019BF,Chen2025FSD}. It is widely used because it limits list growth to one candidate per bit and avoids metric heuristics. However, two structural drawbacks remain:
\begin{itemize}
    \item \emph{Suboptimal counter-hypotheses:} A flipped candidate is generally not the nearest point
    with the complementary bit value, which biases the LLR magnitudes~\cite{Chen2025FSD}.

    \item \emph{Index$\leftrightarrow$Gray conversion:}
    A hardware detector operates on binary indices, whereas Bit-Flip operates on Gray-coded QAM labels. Each flip therefore introduces conversion logic into the critical path.
\end{itemize}

This paper proposes the \emph{Box-Shift} algorithm, an analytical solution to the MCH problem that improves upon Bit-Flip in both computational complexity and coded performance. For each bit, Box-Shift \emph{shifts the existing box} to the nearest node with the complementary bit value, which is the theoretical partially optimal counter-hypothesis for that bit. A closed-form rule identifies this node using one inverter per bit, independently of the QAM order, and its known position simplifies the metric update to an increment. The box also provides a second counter-hypothesis per bit at no additional search cost. On Box Decoding with deterministic pruning~\cite{box_scp}, Box-Shift gains $0.7$--$1.5$\,dB over Bit-Flip and $1.4$--$1.7$\,dB over $K$-Best with Bit-Flip at a bit error rate (BER) of $10^{-5}$ and equal soft-candidate budgets. Against Bit-Flip's explicit metric evaluation, its soft-output stage needs a quarter of the multiplications and one half to two thirds of the operations.\par 

\textbf{Notation:} Scalars, vectors, and matrices are denoted by $a$, $\mathbf{a}$, and $\mathbf{A}$, respectively. The superscript $(\cdot)^H$ denotes the Hermitian transpose, while $|\cdot|$, $\|\cdot\|^2$, and $\lfloor\cdot\rfloor$ denote the absolute value, squared Euclidean norm, and floor operation, respectively. The symbols $\mathbb{C}$, $\Re\{\cdot\}$, and $\Im\{\cdot\}$ denote the complex field and the real and imaginary part operators, respectively. Finally, $\oplus$ denotes the bitwise exclusive-OR (XOR) operation.

%=============================================================================
\section{System Model}\label{sec:system}
%=============================================================================

Consider a spatially multiplexed MIMO system with $N$ transmit and $N$ receive antennas employing bit-interleaved coded modulation~\cite{Caire1998BICM}. Information bits are encoded using an outer code of rate $R$, interleaved, and mapped onto Gray-coded square QAM symbols drawn from the alphabet $\A$, with $|\A|=2^L$. Let $\gd_{i,l}\in\{0,1\}$ denote the $l$-th Gray-coded bit of the $i$-th transmitted symbol. Its bipolar representation is $c_{i,l}=1-2\gd_{i,l}\in\{+1,-1\}$, where $i=1,\ldots,N$ and $l=0,\ldots,L-1$. The received signal is modeled as
\begin{equation}\label{eq:channel}
    \mathbf{y}=\mathbf{H}\mathbf{x}+\mathbf{w},
\end{equation}
where $\mathbf{x}\in\A^{N}$ is the transmitted symbol vector, $\mathbf{H}\in\mathbb{C}^{N\times N}$ is the channel matrix, assumed perfectly known at the receiver, and $\mathbf{w}\sim\mathcal{CN}(\mathbf{0},N_0\mathbf{I}_N)$ is the additive noise vector. For tree-search detection, the sorted QR decomposition (SQRD)~\cite{Wubben2001SQRD} gives $\mathbf{H}=\mathbf{Q}\mathbf{R}$ with $\mathbf{Q}$ unitary and $\mathbf{R}$ upper triangular, its column reordering setting the detection order. The equivalent received signal is then expressed as
\begin{equation}
\mathbf{y}'=\mathbf{Q}^{H}\mathbf{y}
=\mathbf{R}\mathbf{x}+\mathbf{w}',
\end{equation}
where $\mathbf{w}'=\mathbf{Q}^{H}\mathbf{w}$ retains the original noise distribution due to the unitary transformation.\par

The soft-output detector computes the \emph{a posteriori} LLR of each coded bit as~\cite{Milliner2009SOCA}
\begin{equation}\label{eq:llr-exact}
L(c_{i,l}\mid\mathbf{y})=
\ln\frac{\Pr\{c_{i,l}=+1\mid\mathbf{y}\}}
{\Pr\{c_{i,l}=-1\mid\mathbf{y}\}},
\end{equation}
where the sign determines the bit decision and the magnitude represents its reliability. We assume a non-iterative receiver without a priori feedback from the channel decoder and treat the coded bits as equiprobable and statistically independent at the detector. Under these assumptions and additive white Gaussian noise, the max-log approximation of~\eqref{eq:llr-exact} over a reduced candidate list $\Lset\subset\A^{N}$ is~\cite{Milliner2009SOCA}
\begin{equation}\label{eq:llr}
L(c_{i,l}\mid\mathbf{y})\approx
\max_{\mathbf{x}\in\Lset\cap\Xset_{i,l}^{+1}}\lambda(\mathbf{x})
-\max_{\mathbf{x}\in\Lset\cap\Xset_{i,l}^{-1}}\lambda(\mathbf{x}),
\end{equation}
where $\lambda(\mathbf{x})= -\|\mathbf{y}'-\mathbf{R}\mathbf{x}\|^{2}/N_0$ is the log-likelihood metric, and $\Xset_{i,l}^{\pm1}$ denotes the set of $|\A|^{N}/2$ symbol vectors whose bit at position $(i,l)$ equals $\pm1$. Since exhaustive evaluation over $\A^{N}$ is computationally prohibitive, practical detectors retain only a small candidate list. A finite LLR can be computed directly from~\eqref{eq:llr} only if both $\Lset\cap\Xset_{i,l}^{+1}$ and $\Lset\cap\Xset_{i,l}^{-1}$ are nonempty. If either intersection is empty, the MCH problem occurs at bit position $(i,l)$.

%=============================================================================
\section{Hard-Output Box Decoding with Deterministic Pruning}\label{sec:box}
%=============================================================================
This section reviews the hard-output Box Decoding algorithm with deterministic pruning, which forms the basis of the proposed Box-Shift algorithm. Although named Box Decoding, the original algorithm performs MIMO symbol detection rather than channel decoding. In the uncoded setting, its hard symbol decisions are mapped directly to estimated bits, without a separate channel-decoding stage. In the coded receiver considered here, the proposed soft-output extension instead generates bit-level LLRs for a separate LDPC decoder. To distinguish these operations, we retain Box Decoding as the algorithm’s name, use Box detector for the detection module, and reserve decoder for the LDPC decoder.\par 

The box expansion and pruning rules follow~\cite{box} and~\cite{box_scp}, respectively. Detection proceeds through
layers $i=N,\ldots,1$ of the tree, retaining $K$ surviving paths per layer. Here, $K=B=4$, where $B$ denotes
the number of candidates generated per box. For the $k$-th surviving path with previously detected symbols
$\{\hat{x}_{k,j}\}_{j=i+1}^{N}$, the reference node at layer $i$ is given by 
\begin{equation}\label{eq:refnode}
\hat a_{k,i}=\frac{1}{r_{i,i}}\Bigl(y_i'-\sum_{j=i+1}^{N}
r_{i,j}\,\hat x_{k,j}\Bigr).
\end{equation}
Instead of enumerating the entire constellation $\A$, the \emph{box operation} quantizes the reference node $\hat a_{k,i}$ onto the constellation's integer-index grid and selects the $B=4$ points of the box~\cite{box}. Throughout this paper, $n$ and $p$ denote the real-axis (I) and imaginary-axis (Q) indices, respectively. The QAM levels on each axis are indexed by $\{0,1,\ldots,M\}$, where $M=\sqrt{|\A|}-1$, with uniform spacing $\dq=2$. The corresponding real- and imaginary-axis levels are $\gamma+n\dq$ and $\gamma+p\dq$, respectively, where $\gamma=-M\dq/2$ is the lowest level on either axis. The box associated with path $k$ at layer $i$ is represented by the set of index pairs
\begin{equation}\label{eq:box}
\mathcal{B}_{k,i}
=\{\nz,\,\nz+1\}\times\{\pz,\,\pz+1\}.
\end{equation}
Its lower-left corner $(\nz,\pz)$ is obtained by flooring the zero-forcing reference~\eqref{eq:refnode} onto the index grid~\cite{box}:
\begin{equation}\label{eq:boxop}
\nz=\Bigl\lfloor
\frac{\Re\{\hat a_{k,i}\}-\gamma}{\dq}
\Bigr\rfloor,\qquad
\pz=\Bigl\lfloor
\frac{\Im\{\hat a_{k,i}\}-\gamma}{\dq}
\Bigr\rfloor.
\end{equation}
Both indices are then clipped to $\{0,\ldots,M-1\}$ to keep all four candidates within the constellation. The resulting candidate symbols are $\hat x_{k,i}=\gamma+(\nz+u)\dq +\mathrm{j}\bigl(\gamma+(\pz+v)\dq\bigr)$, where $u,v\in\{0,1\}$. For each candidate, the partial Euclidean distance (PED) is computed recursively as
\begin{equation}\label{eq:ped}
\hat d_{k,i}
=\hat d_{k,i+1}
+\bigl|y_i'-{\textstyle\sum_{j>i}}
r_{i,j}\hat x_{k,j}-r_{i,i}\hat x_{k,i}\bigr|^2.
\end{equation}

All four candidates in a box share the same parent metric. Their relative PED ordering therefore depends only on the displacement of the reference node from the lower-left box corner, $\hat\delta_{k,i}
=\hat a_{k,i} -\bigl(\gamma+\nz\dq+\mathrm{j}(\gamma+\pz\dq)\bigr)$. The deterministic pruning method in~\cite{box_scp} exploits this property through two rules:
\begin{itemize}
    \item \emph{Rule~1}: The sign comparisons
    \begin{equation}\label{eq:rule1}
    \dq-2\Re\{\hat\delta_{k,i}\}\gtrless0,\qquad
    \dq-2\Im\{\hat\delta_{k,i}\}\gtrless0
    \end{equation}
    identify the minimum-PED candidate within the box.

    \item \emph{Rule~2}: Two additional comparisons,
    \begin{equation}\label{eq:rule2}
    \Re\{\hat\delta_{k,i}\}
    -\Im\{\hat\delta_{k,i}\}\gtrless0,\qquad
    \Re\{\hat\delta_{k,i}\}
    +\Im\{\hat\delta_{k,i}\}\gtrless\dq,
    \end{equation}
    rank the remaining three candidates without evaluating
    their Euclidean distances.
\end{itemize}

Applying these rules to the $K$ parallel box clusters yields the sort-free \emph{single-step candidate pruning} (SCP) and \emph{iterative candidate pruning} (ICP) methods~\cite{box_scp}, whose computational cost is independent of $|\A|$. SCP retains the Rule-1 winner from each cluster in one parallel, feed-forward step. ICP first orders the candidates within each cluster using Rule~2 and then iteratively merges the clusters to obtain the $K$ globally best surviving paths.

%=============================================================================
\section{Soft-Output Box Detection: Bit-Flip Limitations and the Box-Shift Algorithm}
\label{sec:bs}
%=============================================================================
The deterministic, sort-free candidate selection makes the Box detector hardware-friendly but leaves only a limited candidate set for soft output. In this section, we address this limitation through a Box-Shift approach designed to improve soft-output detection while retaining the deterministic, sort-free nature of Box detection.\par

\subsection{MCH Problem in Box Detection}
\label{sec:challenges}
A box with $B=4$ spans two \emph{adjacent} Gray-coded QAM levels per axis, and adjacent Gray-coded labels differ in exactly one bit. Hence, for Gray-coded square QAM, only two of the $L$ Gray-coded bits vary inside the box while $L-2$ are constant across all four candidates. At each layer the $K$ retained paths leave an MCH with high probability. A soft-output extension must therefore include explicit counter-hypothesis (CH) candidates at every layer while keeping the deterministic, sort-free and QAM-order-independent datapath of the Box detector.\par

The natural first attempt is to apply Bit-Flip. As in most $K$-Best and fixed-complexity soft detectors \cite{Liu2019BF}, the minimum-PED candidate of each layer, the \emph{partial maximum a posteriori (MAP) estimate}, serves as the flipping target, which we call the \emph{anchored node}, and one soft candidate per Gray-coded bit is created by inverting that bit. This raises two problems.

\begin{figure}[t]
\centering
\subfloat[]{%
  \includegraphics[width=0.394\columnwidth]{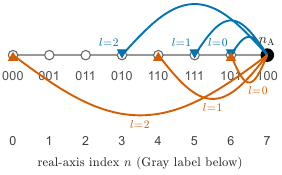}%
  \label{fig:bf}}\hfil
\subfloat[]{%
  \includegraphics[width=0.508\columnwidth]{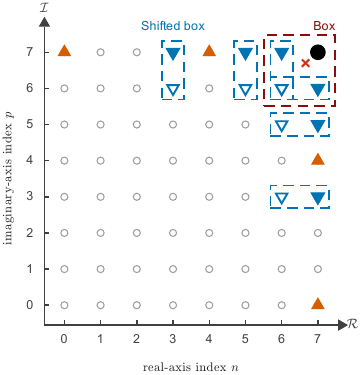}%
  \label{fig:bs}}
\caption{Counter-hypothesis node generation with Box-Shift (blue triangles,
solid: the partially optimal CH node, open: the second CH node) and Bit-Flip
(orange triangles) on (a) the real axis and (b) the whole Gray-coded QAM-64
map, anchored node $(\nA,\pA)=(7,7)$ (black) and reference node (red cross).}
\label{fig:bfbs}
\end{figure}

\subsubsection{Missing the partially optimal CH node}
Flipping bit $l$ of the anchored node's Gray-coded label yields the unique point whose \emph{other} $L-1$ bits agree with it. In a Gray-coded constellation, this point is in general \emph{not} the point nearest to the reference node~\eqref{eq:refnode} that carries the complementary bit value. We call that nearest point the \emph{partially optimal CH node}: among all CH paths at this layer it provides the smallest PED increment and is the most likely maximizer in~\eqref{eq:llr}. Bit-Flip usually lands beyond it, in Fig.~\ref{fig:bf} at the real-axis bits $l=1,2$ with a gap that widens with $l$, which could overestimate the LLR magnitude. Chen et al.~\cite{Chen2025FSD} report the same effect, ``optimistic LLR estimates'' from distant Bit-Flip counter-hypotheses that cause decoding failures, and repair it with an extra LLR-refinement stage. The position of the flipped node is also not known in advance. It is available only after the label has been mapped, flipped and demapped, and its distance from the anchored node varies with the label.

\subsubsection{Index$\leftrightarrow$Gray (de)mapping}
In hardware the box candidate symbol is carried as a binary axis index~\eqref{eq:boxop}, whereas Bit-Flip is defined on the Gray-coded label. Every flip therefore requires a binary-to-Gray mapping of the anchored node and a Gray-to-binary demapping of the result, both on the critical path to the soft candidate's metric. $K$-Best detectors with Bit-Flip carry the same overhead, implemented either as serial Gray-decoding chains \cite{Chen2013FSDVLSI} or as look-up tables that grow with $|\A|$ \cite{Liu2019BF}.

\subsection{Box-Shift}\label{sec:bs-alg}

The Box detector works on the integer index map, and Box-Shift is developed in the index domain. For every Gray-coded bit of the anchored node, Box-Shift shifts the box along that bit's axis onto the \emph{partially optimal CH index} starting from the anchored node. As in most $K$-Best and fixed-complexity soft detectors \cite{Liu2019BF}, the anchored node at layer $i$ is the minimum-PED candidate. The regularity of Gray-coded axes has long been used by soft \emph{demappers} \cite{Tosato2002Demapper} to \emph{compute LLR magnitudes}, whereas here it \emph{locates CH candidates}.

\begin{figure}[t]
\centering
\subfloat[]{%
  \includegraphics[width=0.468\columnwidth]{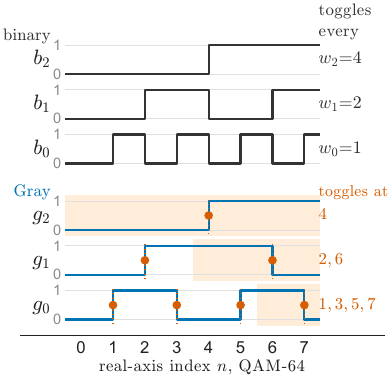}%
  \label{fig:gray}}\hfil
\subfloat[]{%
  \includegraphics[width=0.463\columnwidth]{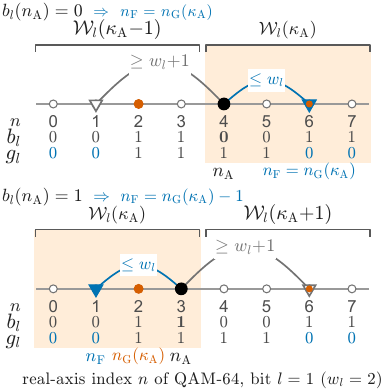}%
  \label{fig:chwin}}
\caption{(a) Binary and Gray digits on one QAM-64 axis with the Gray
crossings (orange dots) and the Gray crossing windows of
$\nA=7$ (shaded); (b) $\nF$ for $\nA$ in the first (top) and the second (bottom) half of its window.}
\label{fig:graych}
\end{figure}
\subsubsection{Closed-form determination of the partially optimal CH index}
The Gray-coded label of a square-QAM symbol is formed independently along the I and Q axes, with $\beta=L/2$ bits per axis. It is therefore sufficient to derive the CH index for one axis. We consider the real-axis index $n$ and the same derivation applies to the imaginary-axis index $p$.\par
For $n\in\{0,\ldots,M\}$, let
$\mathbf{b}(n)=\bigl(b_{\beta-1}(n),\ldots,b_0(n)\bigr)$ denote the binary representation of $n$, i.e., $n=\sum_{j=0}^{\bax-1}\bb_j(n)\,2^{j}$ with $\bb_j(n)\in\{0,1\}$, and let $\mathbf{g}(n)$ denote the corresponding Gray code~\cite{Knuth2011}, with
\begin{equation}
g_l(n)
=
b_l(n)\oplus b_{l+1}(n),
\qquad
b_\beta(n)\equiv 0 .
\label{eq:gray_code}
\end{equation}
For a given bit position $l$, the Gray-coded axis is partitioned into contiguous regions in which $g_l(n)$ has the same value. Transitions between neighboring regions define the decision boundaries for bit $l$ used in soft demapping. Let $\nA$ be the I-axis index of the anchored node. A valid CH index for bit $l$ must lie in a region with the opposite bit value, i.e.,
\begin{equation}
g_l(n_{\mathrm{CH}})
\neq
g_l(\nA).
\label{eq:ch_bit_condition}
\end{equation}
The \emph{partially optimal CH index} $\nF$ is the index nearest to $\nA$ that satisfies \eqref{eq:ch_bit_condition}, which is also the first index beyond the decision boundary closest to $\nA$. We now derive it in closed form.\par

Bit $l$ of the binary representation of $n$ is
\begin{equation}\label{eq:bl}
\bb_l(n)=\Bigl\lfloor\frac{n}{w_l}\Bigr\rfloor\bmod2,
\end{equation}
where $w_l=2^{l}$. $\lfloor n/w_l\rfloor$ increments at every multiple of $w_l$, which means $\bb_l$ flips exactly at the multiples of $w_l$ and is constant in between. Replacing $l$ by $l+1$ gives $\bb_{l+1}(n)=\lfloor n/2w_l\rfloor\bmod2$, which flips at the even multiples of $w_l$ only. By~\eqref{eq:gray_code}, the exclusive-or flips whenever exactly one of $\bb_l$ and $\bb_{l+1}$ flips, so $\gd_l$ flips exactly at the odd multiples of $w_l$,
\begin{equation}\label{eq:cross}
\nG(\kappa)=(2\kappa-1)\,w_l,\qquad \kappa=1,\dots,2^{\bax-l-1},
\end{equation}
which are the \emph{Gray crossings}, i.e., the first indices after the bit-$l$ decision boundaries (Fig.~\ref{fig:gray}). Consecutive crossings are $2w_l$ apart, so $\gd_l$ keeps its value on the run of $2w_l$ indices between them. Around each crossing we define the \emph{Gray crossing window} as the crossing itself, the $w_l$ indices before it and the $w_l-1$ after it,
\begin{equation}\label{eq:window}
\begin{aligned}
\mathcal{W}_l(\kappa)&=\{\nG(\kappa)-w_l,\dots,\nG(\kappa)+w_l-1\}\\
&=\{2(\kappa-1)w_l,\dots,2\kappa w_l-1\},
\end{aligned}
\end{equation}
which contains $2w_l$ indices and exactly one crossing (shaded in Fig.~\ref{fig:gray}). \par
Let $\kA=\lfloor\nA/2w_l\rfloor+1$ be the window of $\nA$, which places $\nA$ at most $w_l$ from its crossing $\nK$. Since $\gd_l$ changes only at crossings, the nearest complementary index to $\nA$ lies either in the window of $\nA$ or in an adjacent window. In the first half of the window (Fig.~\ref{fig:chwin}, top), before the crossing ($2(\kA-1)\,w_l\leq\nA<(2\kA-1)\,w_l$), $\lfloor\nA/w_l\rfloor=2(\kA-1)$ is even by~\eqref{eq:bl}, i.e., $\bb_l(\nA)=0$. The nearest complementary indices are $\nK$ above $\nA$ in the same window, and $\nK-2w_l-1$ below it in the previous window, at the distances
\begin{equation}\label{eq:nearfirst}
\nK-\nA\;\leq\;w_l\;<\;\nA-(\nK-2w_l-1),
\end{equation}
so the partially optimal CH index is $\nK$.\par
In the second half (Fig.~\ref{fig:chwin}, bottom), from the crossing on ($(2\kA-1)\,w_l\leq\nA\leq2\kA w_l-1$), $\lfloor\nA/w_l\rfloor=2\kA-1$ is odd, i.e., $\bb_l(\nA)=1$. The nearest complementary indices are $\nK-1$ below $\nA$ in the same window, and $\nK+2w_l$ above it in the next window, at the distances
\begin{equation}\label{eq:nearsecond}
\nA-(\nK-1)\;\leq\;w_l\;<\;(\nK+2w_l)-\nA,
\end{equation}
so the partially optimal CH index is $\nK-1$. In both cases it is nearer by at least one index, while the reference $\hat a_{k,i}$ lies within half an index of $\nA$. It is therefore also the partially optimal CH node of bit $l$. With
\begin{equation}\label{eq:nkappa}
\nK=(2\kA-1)\,w_l=2^{\,l+1}\Bigl\lfloor\frac{\nA}{2^{\,l+1}}\Bigr\rfloor+2^{\,l},
\end{equation}
the partially optimal CH index is
\begin{equation}\label{eq:nf}
\nF=\nK-\bb_l(\nA).
\end{equation}
Since \eqref{eq:nkappa} is a mask-and-set operation on $\nA$ and \eqref{eq:bl} is a bit selection, the binary representation of $\nF$ in \eqref{eq:nf} is
\begin{equation}\label{eq:onenot}
\bb_j(\nF)=
\begin{cases}
\bb_j(\nA), & j>l,\\[1pt]
\overline{\bb_l(\nA)}, & j=l,\\[1pt]
\bb_l(\nA), & j<l,
\end{cases}
\end{equation}
i.e., $\nF$ inverts bit $l$ of $\nA$, keeps the bits above it and fills the bits below it with $\bb_l(\nA)$. Over all $\bax$ axis bits the index set $\{\nA\}\cup\{\nF^{(l)}\}_{l=0}^{\bax-1}$ covers both values of every axis bit. Since the I and Q Gray-coded labels are independent, the box and the Box-Shift soft nodes together are MCH-free at all $L$ bits of the layer.

\subsubsection{The two Box-Shift mechanisms}
For an I-axis bit, Box-Shift shifts the box along the I axis onto $\nF$, which gives the \emph{shifted box} $\{\nF\}\times\{\pz,\pz+1\}$ (dashed in Fig.~\ref{fig:bs}). For a Q-axis bit, the shifted box is $\{\nz,\nz+1\}\times\{\pF\}$. Both nodes of the shifted box carry the complementary bit value. The \emph{closest-node mechanism} (Box-Shift-CN) keeps only the nearer one, the partially optimal CH node $(\nF,\pA)$ or $(\nA,\pF)$ (solid triangles), giving $L$ soft candidates per symbol. The \emph{full-box mechanism} (Box-Shift-FB) also keeps the second CH node one $\dq$ step away (open triangles), giving $2L$ candidates per symbol without any further search.\par
Each CH node starts a protected path that inherits the anchored path's history, bypasses the pruning and is extended at every lower layer by its Rule-1 winner, in the same feed-forward pass as the hard-decision paths. The LLR~\eqref{eq:llr} is then computed over the $K+NL$ leaves of Box-Shift-CN, or the $K+2NL$ leaves of Box-Shift-FB.

\begin{figure*}[t]
\centering
\includegraphics[width=\textwidth]{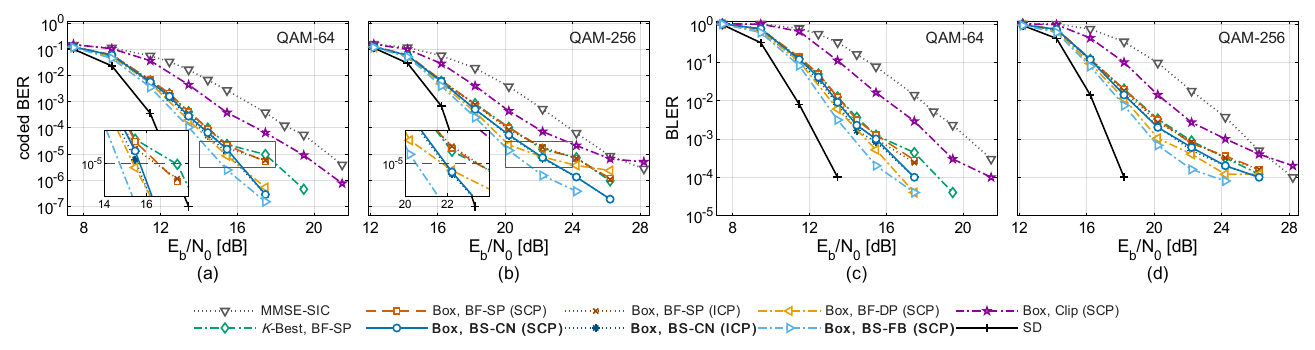}
\caption{Coded BER (a), (b) and BLER (c), (d) of Box-Shift (BS) and
Bit-Flip (BF) versus $E_b/N_0$ at QAM-64 and QAM-256: $4\times4$ MIMO-OFDM
over TDL-C with 5G~NR LDPC, $R=3/4$.}
\label{fig:ber} 
\end{figure*}

%=============================================================================
\section{Numerical Evaluation}\label{sec:results}
%=============================================================================

This section evaluates the computational complexity and coded error performance of Box-Shift against Bit-Flip. To isolate the effect of counter-hypothesis generation, both methods use the same Box detector with deterministic pruning and $K=B=4$. Comparisons consider equal soft-candidate budgets per symbol: $L$ for Box-Shift-CN and Bit-Flip-SP, and $2L$ for Box-Shift-FB and Bit-Flip-DP. Bit-Flip-SP operates on the minimum-PED candidate at each layer, whereas Bit-Flip-DP additionally uses the second-best candidate.\par

\subsection{Complexity Comparison}\label{sec:complexity}

The soft-output stage in each layer includes \emph{finding the CH candidates} and the corresponding \emph{PED computation}.

\emph{\textbf{Finding the CH candidates:}} Box-Shift needs only the single inverter of \eqref{eq:onenot} per bit, independent of the QAM order and without any Gray-code conversion. Every Bit-Flip flip, in contrast, passes through the Gray (de)mapping, realized as serial decoding chains or as look-up tables that grow with $|\A|$ \cite{Liu2019BF}. The second counter-hypothesis per bit needs no search in Box-Shift-FB, whereas Bit-Flip-DP needs a second-minimum search per layer.

\emph{\textbf{PED computation:}} Table~\ref{tab:cost} lists the per-symbol PED computation complexity. A Box-Shift soft node differs from the anchored node, and the second Box-Shift-FB node from the partially optimal CH node, by a known integer offset on one axis. With $\Delta n=\nF-\nA$ and the real-axis displacement of the reference node from $\nA$ in units of $\dq$, $\varepsilon=(\Re\{\hat a_{k,i}\}-\gamma)/\dq-\nA$, \eqref{eq:ped} expands to
\begin{equation}\label{eq:pedoffset}
\hat d(\nF)=\hat d(\nA)+r_{i,i}^{2}\dq^{2}\,\Delta n\,(\Delta n-2\varepsilon).
\end{equation}
For the CH node of each bit, the update costs one real multiplication by the per-layer constant $r_{i,i}^{2}\dq^{2}$ and four real additions, namely $\Delta n=\nF-\nA$, the subtraction in the parentheses, the product with $|\Delta n|\leq 2^{\bax-1}$, which reduces to shifts and at most one addition for QAM-256 and below, and the accumulation. The displacement $\varepsilon$ costs one subtraction per layer, i.e., per symbol, since $(\Re\{\hat a_{k,i}\}-\gamma)/\dq$ comes from \eqref{eq:boxop}, and the Box-Shift-FB partner node with $|\Delta n|=1$ costs one multiplication and two additions per bit. The position of a Bit-Flip node is known only after the Gray mapping, flip and demapping, so its offset from the anchored node is not available when the increment of \eqref{eq:pedoffset} is formed. Table~\ref{tab:cost} therefore counts the Bit-Flip node with the explicit calculation of the accumulated PED~\eqref{eq:ped} as in its reference implementation~\cite{box_scp,Chen2013FSDVLSI}, four multiplications and four additions. An incremental Bit-Flip update is possible once the offset is known, i.e., after the Gray (de)mapping. Under this convention Box-Shift spends a quarter of Bit-Flip's multiplications, one more addition per symbol with $L$ candidates and nearly a quarter fewer with $2L$.\par

\begin{table}[t]
\caption{Per-symbol PED computational cost of the soft-output stage for an
$L$-bit symbol.}
\label{tab:cost}
\centering
\footnotesize
\setlength{\tabcolsep}{5pt}
\begin{tabular}{@{}lcccl@{}}
\toprule
Scheme & \#CH nodes& \#MUL & \#ADD & Mechanism\\
\midrule
\textbf{Box-Shift-CN} & $L$ & $L$ & $4L+1$ & closest node\\
Bit-Flip-SP & $L$ & $4L$ & $4L$ & single path\\
\textbf{Box-Shift-FB} & $2L$ & $2L$ & $6L+1$ & full box\\
Bit-Flip-DP & $2L$ & $8L$ & $8L$ & double path\\
\bottomrule
\end{tabular}
\end{table}

\subsection{Coded BER Performance}\label{sec:ber}

Link-level simulations use a $4\times4$ MIMO-OFDM link over the 3GPP TDL-C channel (delay spread $300$\,ns, independent fading realization per code block) with 5G~NR LDPC coding (base graph 1, $1056$ information bits, $R=3/4$) at QAM-64 and QAM-256. All detectors see the same bits, channel and noise samples and clip LLR magnitudes at $8$. Box-Shift and Bit-Flip (abbreviated as BS and BF in Fig.~\ref{fig:ber}, respectively) run on the identical Box detector with the SCP and ICP pruning rules \cite{box_scp} ($K=B=4$). $K$-Best ($K=4$) with single-path Bit-Flip \cite{Liu2019BF}, $K$-Best with Bit-Flip-SP, is the external baseline at equal surviving nodes per layer, and soft MMSE successive interference cancellation (MMSE-SIC) and SD serve as the BER upper and lower bounds. Box Decoding has been studied only with $B=4$~\cite{box,box_scp}, where it performs close to $K$-Best with $K=4$, so $K=B=4$ isolates the counter-hypothesis generation.\par

On the same Box detector, Box-Shift outperforms Bit-Flip for both candidate budgets (Fig.~\ref{fig:ber}). With the $L$-candidate mechanisms, Box-Shift-CN gains $0.9$ and $1.5$\,dB in $E_b/N_0$ over Bit-Flip-SP at BER~$10^{-5}$ for QAM-64 and QAM-256, respectively, and with the $2L$-candidate mechanisms Box-Shift-FB gains $0.7$ and $1.2$\,dB over Bit-Flip-DP. In both cases the gain grows with the modulation order and toward low BER, because Bit-Flip's flipped node lies farther from the Gray crossing at higher order and overstates the LLRs. The block error rate (BLER) (Fig.~\ref{fig:ber}(c), (d)) shows the same ranking: at BLER~$10^{-3}$, Box-Shift-CN gains $0.3$ and $0.5$\,dB over Bit-Flip-SP and Box-Shift-FB $0.6$ and $0.4$\,dB over Bit-Flip-DP for QAM-64 and QAM-256, respectively. The Box detector with LLR clipping (Box, Clip) trails Box-Shift-CN by $3.7$ and $3.6$\,dB at BER~$10^{-5}$ for QAM-64 and QAM-256, respectively.\par

\begin{figure}[t]
\centering
\includegraphics[width=0.82\columnwidth]{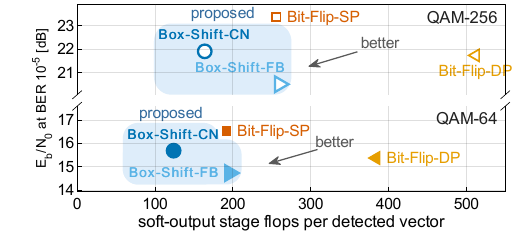}
\caption{$E_b/N_0$ at coded BER $10^{-5}$ against the soft-output stage
flops per detected vector, QAM-64 (filled) and QAM-256 (open).}
\label{fig:tradeoff}
\end{figure}

The ICP curves overlap the SCP curves, so the pruning rule does not affect the coded BER. Against the external baselines, Bit-Flip-SP on the Box detector almost matches $K$-Best with Bit-Flip-SP at equal nodes per layer, and Box-Shift-CN on the Box detector gains $1.4$--$1.7$\,dB over $K$-Best with Bit-Flip-SP at the same candidate budget at BER~$10^{-5}$, while the Box detector with Box-Shift keeps the $4$--$6$\,dB margin over MMSE-SIC of its hard-output form \cite{box_scp}. The gap of the $B=4$ box to the SD bound grows with the QAM order, and a larger box would be needed to reduce it. Fig.~\ref{fig:tradeoff} shows the trade-off between performance and the soft-output stage complexity of Table~\ref{tab:cost}, where Box-Shift-CN and Box-Shift-FB occupy the efficient lower-left corner.

%=============================================================================
\section{Conclusion}\label{sec:conclusion}
%=============================================================================

Box-Shift extends the sort-free, QAM-order-independent Box Decoding to coded systems by solving its missing-counter-hypothesis problem. It locates the partially optimal counter-hypothesis node of every Gray-coded bit in closed form, with one inverter per bit and an incremental metric update. The shifted box supplies a second counter-hypothesis per bit without a further search. Coded $4\times4$ simulations show $0.7$--$1.5$\,dB gains over Bit-Flip at BER $10^{-5}$ at equal candidate budget, a $1.4$--$1.7$\,dB lead over $K$-Best with Bit-Flip-SP at the same candidate budget and a $4$--$6$\,dB margin over MMSE-SIC, at a quarter of the multiplications and one half to two thirds of the soft-output stage operations of Bit-Flip's explicit metric evaluation. Future work includes a VLSI implementation of Box-Shift algorithm and iterative detection--decoding.

\bibliographystyle{IEEEtran}
\bibliography{references}

\end{document}